\documentclass[pdftex,pre,twocolumn,byrevtex,superscriptaddress]{revtex4}

\usepackage{amsmath}
\usepackage{amssymb}
\usepackage{graphicx}
\usepackage{color} 
\newcommand{\req}[1]{(\ref{#1})}

\newcommand{\eps}{\epsilon}

\def\ps{\rho}

\begin{document}

\title{
Let my people go (home) to Spain: a genealogical model of Jewish identities since 1492
}

\author{
  \firstname{Joshua S.}
  \surname{Weitz}
  }
\email{jsweitz@gatech.edu}
\homepage{http://ecotheory.biology.gatech.edu}
\affiliation{
   	School of Biology,
	Georgia Institute of Technology,
	Atlanta, GA 30332
	}
\affiliation{
   	School of Physics
	Georgia Institute of Technology,
	Atlanta, GA 30332
	}

\date{\today}

\begin{abstract}
The Spanish government recently announced an official fast-track path to citizenship for any individual who is Jewish and whose ancestors were expelled from Spain during the inquisition-related dislocation of Spanish Jews in 1492. It would seem that this policy targets a small subset of the global Jewish population, i.e., restricted to individuals who have retained cultural practices associated with ancestral origins in Spain. However, the central contribution of this manuscript is to demonstrate how and why the policy is far more likely to apply to a very large fraction (i.e., the vast majority) of Jews. This claim is supported using a series of genealogical models that include transmissable "identities" and preferential intra-group mating. Model analysis reveals that even when intra-group mating is strong and even if only a small subset of a present-day population retains cultural practices typically associated with that of an ancestral group, it is highly likely that nearly all members of that population have direct geneaological links to that ancestral group, given sufficient number of generations have elapsed. The basis for this conclusion is that not having a link to an ancestral group must be a property of all of an individual's ancestors, the probability of which declines (nearly) superexponentially with each successive generation. These findings highlight unexpected incongruities induced by genealogical dynamics between present-day and ancestral identities.
\end{abstract}

\maketitle

\section*{Introduction}
Present-day Jews predominantly self-identify as either Sephardic or Ashkenazi.
Origins of Sephardic
Jews are generally attributed to the Jewish community based in Spain and Portugal
that was expelled from the Iberian penninsula in the late 15th century,
whereas Ashkenazi Jews generally attribute their origins to Central and Eastern 
Europe, pre-dating the expulsion~\cite{telushkin,zohar_2005}.  These divisions are, at least culturally, considered to be long-standing,
for example, the protagonist of the classic 19th century farce ``The King of Schnorrers'' (which is set in the late 18th century) -- Manasseh Bueno Barzillai Azevedo da Costa --
reacts in horror at the prospect of his daughter marrying an Ashkenazi, rather than
a Sephardic, Jew: ``A Sephardi cannot marry a Tedesco [Ashkenazi]! It would be a degradation''~\cite{zangwill}.
The conditions for the fast-track naturalization 
announced by the Spanish
government in November 2012, follow along these traditional
designations of ethnic identity: (i) the petitioner must be Jewish;
(ii) the petitioner must ``certify'' their Spanish Jewish origins.
Indeed, the announcement and subsequent media coverage
of this change to Spanish civil law highlighted
its intended target to be self-identified Sephardic Jews 
~\cite{sjews_nytimes,sjews_elpais} --
estimated to comprise 20\% of the global Jewish population.
Here, the following question is asked: to what extent should
\emph{any} Jew living today expect to
have one (or more) Jewish ancestors expelled from Spain in 1492.

Although identities may indeed be exclusive and even strongly retained
inter-generationally,
this does not preclude the fact that individuals of one identity may have one
(or more) ancestors of a different identity.    
Irrespective of identity, the number of ancestors that any individual
has grows quite rapidly, exponentially at first and then increasing (albeit
at a slower-than-exponential rate) 
with each successive prior generation.   Hence, a present-day individual that
self-identifies as Jewish would have
a direct genealogical link to the expelled Spanish Jewish community 
if one (or more)
of their deceased forebears in 1492 was a member of that community. 
Here, the main contribution is to develop a simple genealogical model
(and intuition) to explain how having ancestors of diverse
``types'' is extremely common even when cross-``type'' mating is rare.

\section*{Model of genealogical dynamics with assortative mating}
Consider the genealogical dynamics of
a population of $N(g)$ individuals where $g=0$ denotes
the ancestral population of interest (e.g., the number of Jews living
in Spain in 1492) and $g>0$ denotes each successive prior generation such
that $g=g_0$ denotes the present.  
In this model,
given an initial population size, $N(0)$, then
$N(1)$ pairs of parents are selected at random from the $N(0)$ individuals
present in the $0$-th generation, forming a history of genealogical
links between each generation.  The value of $g$ is then incremented iteratively
from $1$ to $g_0$. In this model, individuals can be selected more than
once and no information on male/female identities are retained.  
These dynamics correspond to the forward version of standard models
of genealogical dynamics with sexual 
reproduction~\cite{chang_1999,derrida_2000}.
Statistical properties of genealogical dynamics at the population scale have
previously been found to be highly robust to this apparent lack of 
realism~\cite{manrubia_2003,rohde_nat2004,brunet_2013}. 

\begin{figure}
\begin{center}
\includegraphics[width=0.45\textwidth]{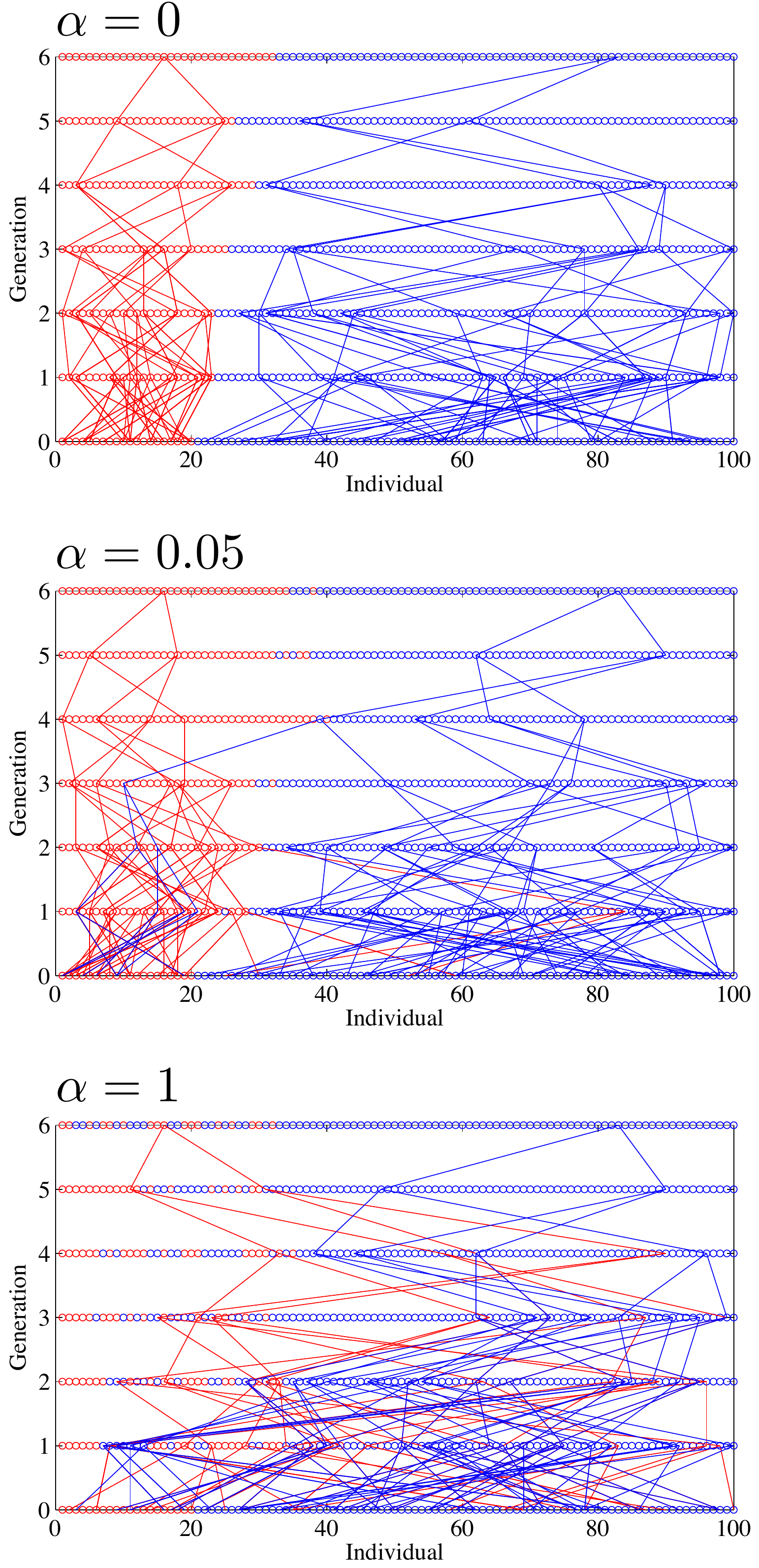}
\end{center}
\caption{Genealogical history with mating preferences.  Each
panel illustrates the ancestors of two focal individuals given $N=100$
and $\alpha=0, 0.05 \mbox{\textrm{ and }} 1$ (top-middle-bottom). Initially,
at $g=0$, there are 20\% type-1 individuals (red circles) and
80\% type-2 individuals (blue circles).  Ancestors are denoted
by red and blue lines, respectively.   When viewing the process 
retrospectively from the present-day (top line in each panel, $g=6$),
it is apparent that all ancestors share the same identity as the focal
two individuals when $\alpha=0$.  However, when $\alpha> 0$ then
one (or more) ancestors may have a different identity than that of the
focal individual.
\label{fig.mixing}}
\end{figure}

Next, consider a modified version of the standard model of genealogical dynamics with sexual reproduction
to distinguish between individuals
with two exclusive traits, type $1$ or type $2$, e.g., 
Sephardic (type-1) or Ashkenazi (type-2) These traits need not be genetically encoded. 
In this framework, each generation can be described in
terms of a population-wide distribution of traits, $\mathbf{x}={x_1,x_2,\ldots,
x_N}$ where $x_i=\{1,2\}$.   The fraction of type-1 individuals is
denoted as $p(g)$.
Individuals of different traits may prefer to mate with individuals
with the same trait.  A generalized
mating preference parameter $\alpha$ can account for preferential
mating, such that
the $N(g)$ 
parents of individuals in generation $g-1$ will be selected
from a multinomial distribution with probabilities:
$P(11)=p\left[(1-\alpha)+p\alpha\right]$,
$P(12 \textrm{ or } 21) =2p\left[\alpha(1-p)\right]$,
and $P(22)=(1-p)(1-\alpha p)$,
where $p$ is the proportion of type-1 individuals in generation $g-1$.
Hence, when $\alpha=0$, mating occurs exclusively amongst individuals of the
same type, i.e., $P(12 \textrm{ or } 21) = 0$.  When $\alpha=1$, mating
probabilities depend on population proportions exclusively, i.e., $P(11)=p^2$, $P(12 \textrm{ or } 21) = 2p(1-p)$,
and $P(22)=(1-p)^2$. Intermediate values of $0<\alpha<1$ provide a continuum
between these limits, consistent with models of varying degrees
of assortative mating in population genetics.
The model further presumes
that children of two type-1 parents self-identify as type-1 individuals,
and that children of two type-2 parents self-identify
as type-2 individuals. Finally, the model assumes that children of a type-1
and a type-2 parent self-identify as either type-1 or type-2 with
equal probability. 

\section*{Results}

\subsection*{Individuals have ancestors with multiple identities, even when mating is strongly-preferred among individuals of the same identity}
Trait-associated genealogical dynamics were simulated in the modified model,
described above,
for three values of $\alpha=0, 0.05, \mbox{\textrm{ and }} 1$ (see Figure~\ref{fig.mixing}).   When
$\alpha=0$, then the identities of ancestors must match that
of present-day individuals, since there is no reproduction
among individuals of different identities.  In this limit,
the fraction of present-day individuals with no type-1 ancestors 
remains relatively constant (near $1-p_0$).  Whereas,
when $\alpha=1$, then reproduction occurs irrespective of identity.  Hence, 
it is highly likely
that individuals have ancestors of both identities. Notably,
the same phenomenon is observed
even when preferences are strong for intra-type reproduction, e.g.,
when $\alpha=0.05$.  The reason
is occasional (even rare) instances of cross-identity
reproduction among an exponentially growing number of ancestors
(and matings) lead to frequent instances of type-1 individuals with
type-2 ancestors and vice-versa.  Moreover, once a type-1 individual has
at least one type-2 ancestor, then the mating preferences reinforce this
history, ensuring that type-1 individuals often have many type-2 ancestors
(and vice versa).
These exploratory simulations reveal that 
individuals can have ancestors with identities other than their
own, even when individuals strongly prefer to reproduce wth individuals
of the same identity.  

\subsection*{The likelihood of having an ancestor with an identity different than one's own increases (nearly) super-exponentially with prior generations}
\begin{figure}
\begin{center}
\includegraphics[width=0.45\textwidth]{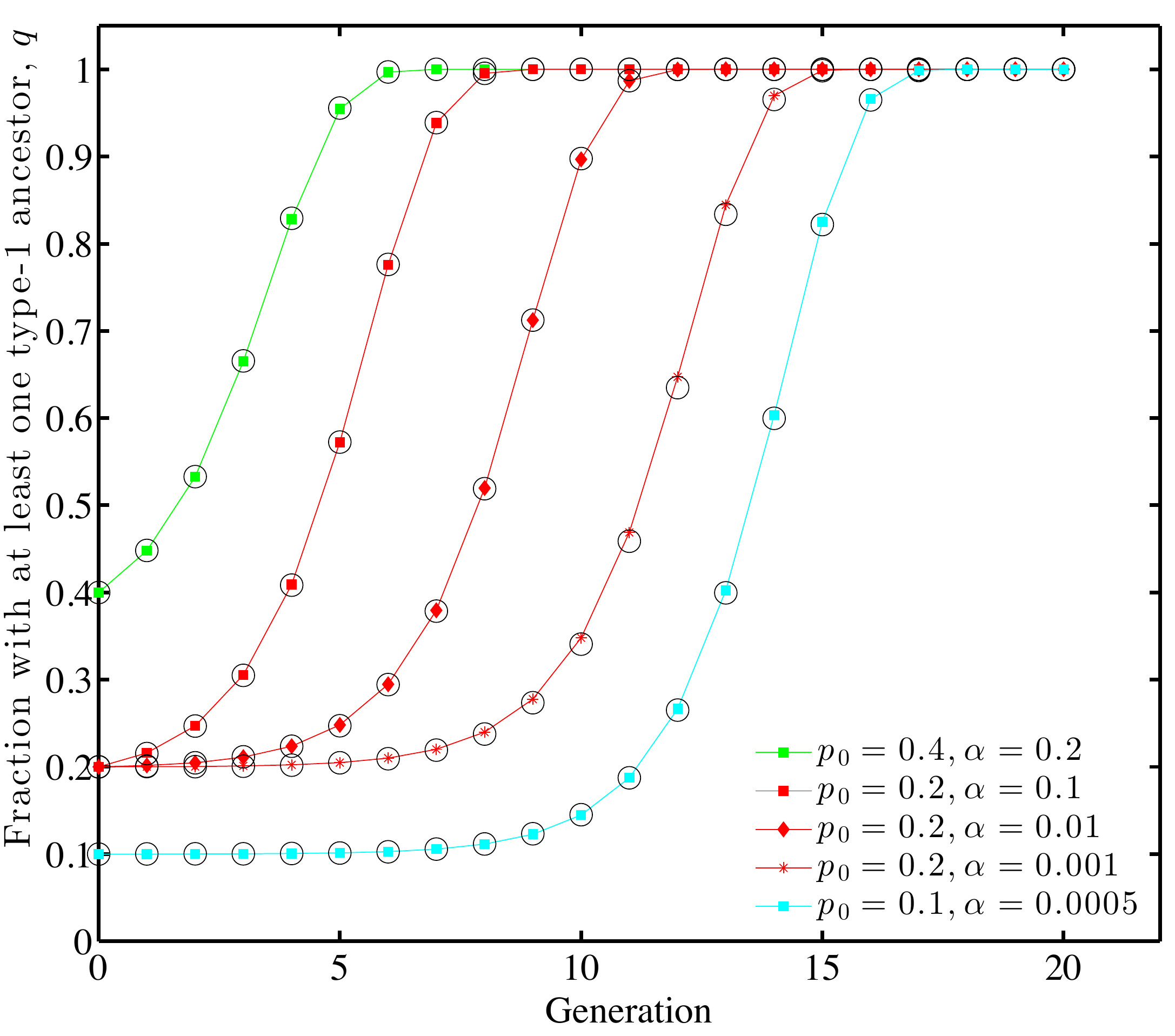}
\caption{Rapid increase in frequency of genealogical link to a previously minority
identity.  Solid lines with symbols 
denote theory of Eq.~\req{q} 
(colored based on initial frequency of type-1 individuals) while circles denote
simulation. The simulated population has $N=10^6$ individuals.
\label{fig.superexp}}
\end{center}
\end{figure}

Introducing the concept of \emph{genealogical identity} 
will prove useful to
explore the phenomenon of 
having ancestors with identities different than that of a focal individual,
Let
$y_i=\{0,1\}$ represent whether or not any of a focal $i$-th individual's
ancestors had a particular identity of interest.  In this study system,
$y=0$ denotes that none of an individual's ancestors was of type-1,
whereas $y=1$ denotes that at least one ancestor of a focal individual
was of type-1.  Hence, the state of an individual can be described in
terms of $s_i=(x_i,y_i)$.  Due to the definition of identity transfer
in the model, there are three possible states of individuals:
(i) $s=(1,1)$ - a type-1 individual with at least one type-1 ancestor;
(ii) $s=(2,0)$ - a type-2 individual with no type-1 ancestors;
(iii) $s=(2,1)$ - a type-2 individual with at least one type-1 ancestor.
Analysis of genealogical dynamics yield a closed form
prediction for the fraction of individuals with at least one type-1 ancestor:
\begin{equation}
q(g) = 1 - \left(1-p_0\right)\left(1-\alpha p_0\right)^{2^{g}-1},
\label{q}
\end{equation}
(see derivation in the Appendix).
The
fraction of individuals with at least one type-1 ancestor
rapidly approaches one, given that
the probability of having no type-1 ancestors declines (nearly) super-exponentially.
This prediction can be evaluated for different values of
$\alpha$ and $p_0$ in a large population simulation where $N=10^6$.
Quantitative agreement is found even when $\alpha\ll 1$ (see Figure~\ref{fig.superexp}).
Indeed, even when the vast majority
of sexual reproductions occur between individuals of the same identity 
as occurs when $\alpha\ll 1$, the fraction of individuals with 
at least one type-1 ancestor, $q$, converges rapidly from $q_0<0.5$
to $q_0=1$.

\begin{figure}
\begin{center}
\includegraphics[width=0.45\textwidth]{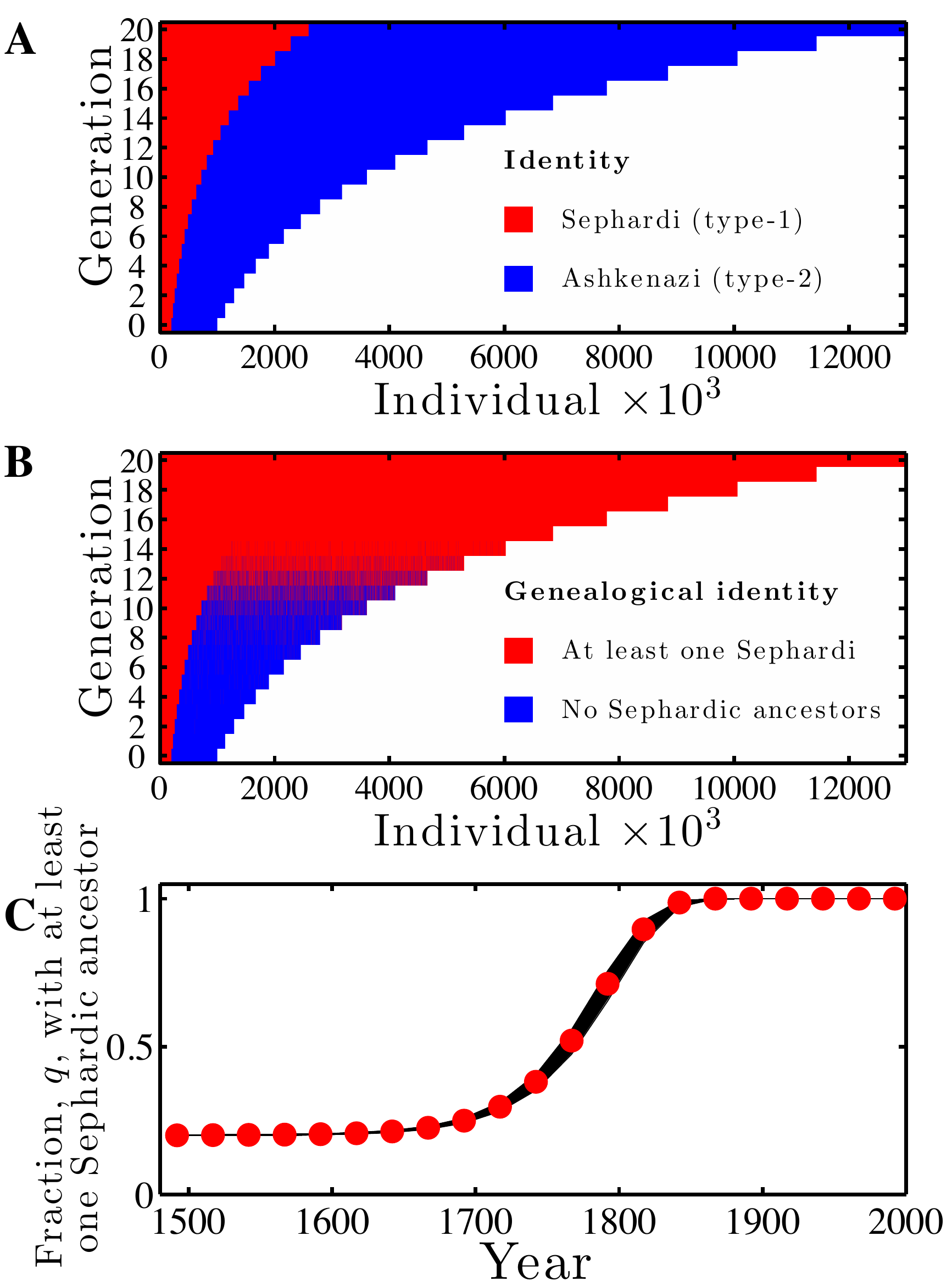}
\caption{Nearly all modern-day Jews are likely to have many ancestors expelled 
from Spain in 1492.  Geneaological dynamics are simulated using $\alpha=0.00124875$
in a population of initial size $1\times 10^6$ at generation 0 that
is of size $1.3\times 10^7$ at generation 20.  The choice of $\alpha$ corresponds
to a relative preference of in-group mating of 1000:1 relative to out-group mating.
(A) Heat-map of identities, $x(g)$.  (B) Heat-map of genealogical identities, $y(g)$.
(C) Comparison of theoretical prediction (red circles) of the fraction of individuals
with at least one Sephardic ancestor, $q$, with $10^3$ simulations (all variation contained in black shaded region).
\label{fig.applied}}
\end{center}
\end{figure}

\subsection*{Nearly all present-day Jews are likely to have at least one (if not many more) ancestors expelled from Spain in 1492}
The modified genealogical model with mating preferences can be
evaluated in a
parameter regime inspired by that of transmission of Sephardic and
Ashkenazi identities.  This regime includes the assumptions
that $N(0)=1\times 10^6$ and
that $N(20)=1.3\times 10^7$, spanning an approximately 500 year time frame,
corresponding to approximately 20 generations~\cite{botticini_2007,dellapergola_2012}.
Further, the application of the model assumes that the Sephardic
community was approximately 20\% of the global Jewish population in 1492.
Mating preference data spanning this historical period
is not available in a comprehensive fashion.  
Instead, consider a highly conservative (and highly biased)
mating scenario where $\alpha=0.00124875$, corresponding to
a 1000:1 relative likelihood given one type-1 individual to mate with
another type-1 individual rather than a type-2 individual.  Note that surveys
of ``inter-marriage'' between Sephardic and Ashkenazi indvididuals in 
modern Israel
suggest that within-ethnicity marriages account for approximately 90\%
of all marriages~\cite{rubinstein_2011}.  Such modern estimates
suggest that strong inter-type preferences continue to have persisted
into the 20th century.  
Simulations
reveal that, despite an extreme preference for in-group mating, it takes only
15 generations for all individuals to have at least one direct genealogical
link to a Sephardic in the $g=0$ generation (see Figure 3B).  This occurs despite
the fact that Sephardic identity remains a minority throughout the simulation
(see Figure 3A).  The frequency of individuals with at least one type-1 
ancestor, $q(g)$, agrees with theoretical predictions (see Figure 3C), further
substantiating the generality of the present 
mechanism for the spread of genealogical
identity in constant and in varying populations.
These results are highly robust to changes in initial population fractions,
$p_0$, population sizes $N(g=0)$ and $N(g=20)$, and mating preferences $\alpha$.

\begin{figure}
\begin{center}
\includegraphics[width=0.45\textwidth]{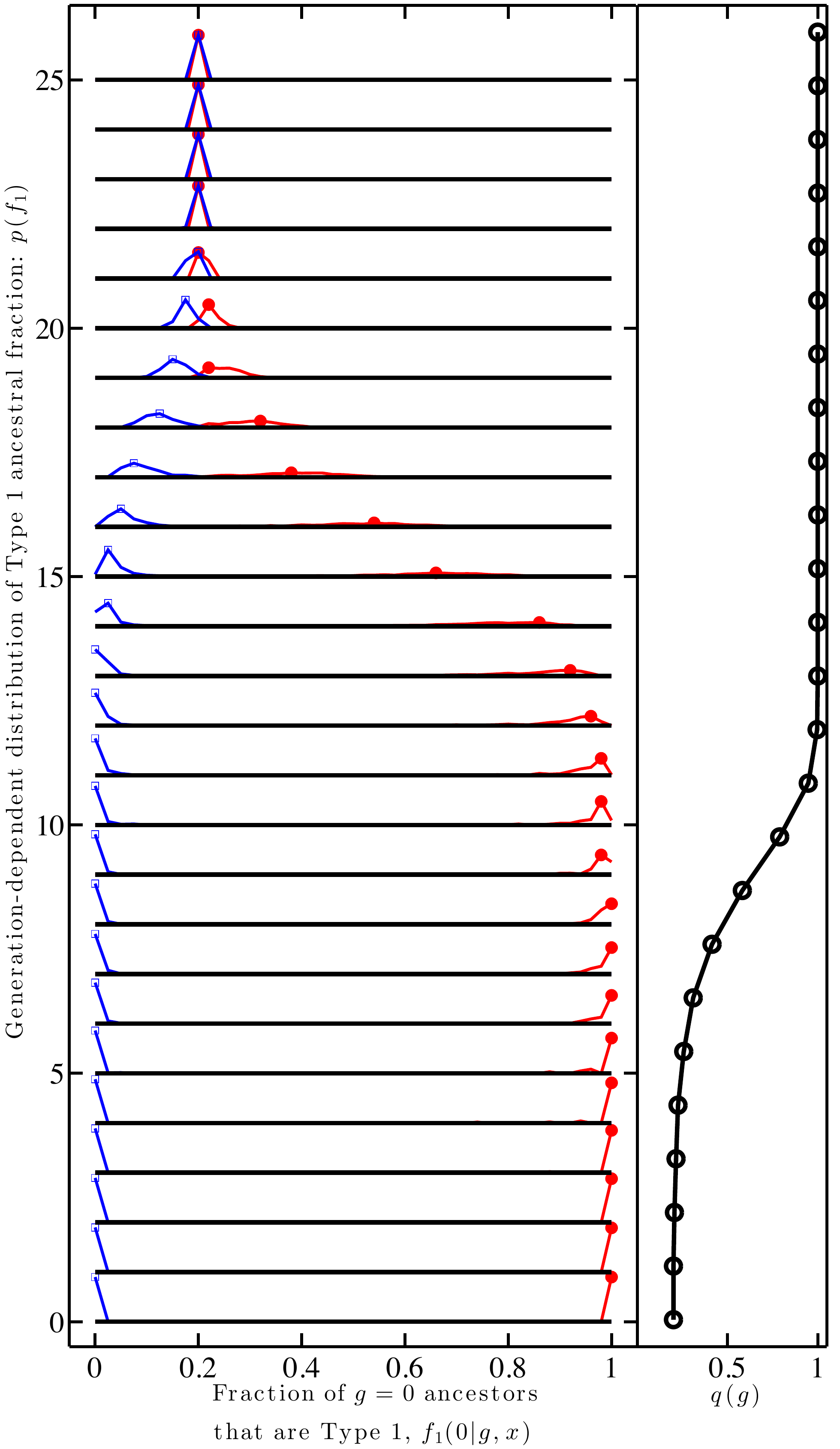}
\caption{Individuals have many ancestors with different identities than
their own.  The results are from a simulation of a population with
$N=10^4$, $g=25$, $p_0=0.2$ and $\alpha=0.01$.  (Left) Fraction of ancestors in $g=0$ that are type 1,
for individuals who at generation $g$ self-identify as type 1 (red)
or type 2 (blue); (Right) Fraction of individuals with
at least one type-1 ancestor, $q(g)$. Generations increase from $g=0$
to $g=25$ along the y-axis in both panels. \label{fig.many}}
\end{center}
\end{figure}

Claims regarding the diversity of ancestral identities can be extended
using the same model framework.
Specifically, the model predicts that
individuals are not only likely to have at least one ancestor
of an identity different that their own (as shown in Figures 2 and 3), 
but are likely to have \emph{many}
such ancestors.  To illustrate this concept, consider a 
type-1 individual in generation $g_0$ who has one type-2 ancestor in generation $g_0-k$.  Then, the type-1 individual
would be a direct genealogically descendant of many type-2 ancestors in
generation $0$ via this (rare) link, subsequently reinforced by assortative 
mating. A quantitative metric can be introduced
to characterize this phenomenon: $f_1(0|g,x)$ - the fraction of ancestors in generation
0 who are of type 1 for an individual living in generation $g$ whose
identity is $x$.  Simulations reveal 
that $f_1(0|g,x=1)$ converges to that of $f_1(0|g,x=2)$ over time.  This implies that individuals of different identities both
share many, and eventually, all ancestors of a focal identity (see Figure~\ref{fig.many}).  
This claim is
consistent with prior analysis of panmictic populations
in which there is a rapid (i.e., scaling independently of
population size) and recent (i.e., scaling with the logarithm 
of poulation size) transition, in terms of generations,
over which two
randomly chosen individuals are likely to switch from sharing
very few to nearly all of their
ancestors~\cite{derrida_2000}.

\section*{Discussion}
In summary, a series of simplified genealogical models have been proposed and analyzed
in which individuals retain a set of identities.  
Nearly all model variants lead to dynamics in which present-day individuals who identify
with one identity have at least one, and typically many, ancestors of a different
identity in a relatively recent
generation.
As is apparent, the model formulation is generic rather than specifically parameterized
for the detailed and complex structure of present-day and historical
Jewish populations - which inspired the present analysis.
Hence, extensions are warranted that include population structure,
mating with other ``groups'' 
(e.g., those who do not transmit or retain
the cultural practices associated with the identity), loss of identity, 
and other demographic structure that may change both the baseline expectation
and variation for identity dynamics as described by the present model.
Nonetheless, given the super-exponential nature of the process, 
it would require significant changes to the model structure to substantively
modify the overall conclusion: given occasional mating between
individuals of different identities, then a small
number of generations is likely required for individuals to have a subset of their
ancestors with identities different than their own.
This notion is strongly consistent with landmark work on the genealogical
ancestry of all living humans~\cite{rohde_nat2004}
and population genomics research
on the related ancestry of individuals from seemingly disparate European 
populations~\cite{ralph_plos2013}.

Returning to the inspiration for this model, note that
the Spanish government's fast track to citizenship 
was ostensibly meant to
target Sephardic Jews, a minority of the global Jewish population.  
Extrapolating from the model analysis, it would seem that, to the contrary,
the far more likely baseline hypothesis is that the vast majority of
present-day Jews have one or more direct genealogical forebears in the Jewish community expelled from
Spain in 1492.  The current analysis does not consider the policy implications
of such a hypothesis. However, it is worth noting that the
fast-track naturalization process announced in November 2012 
has yet to be implemented 
(as of May 2013)~\cite{sjews_many} and that Portugal has also 
recently announced a similar policy modeled on the Spanish framework~\cite{time_portugal}.
Instead, the model can be used to point out that policies linked to 
the identity of (distant) ancestors should be approached cautiously, 
given that the number of
ancestors grows (nearly) exponentially.  As a consequence,
the identity-associated characteristic of ancestors need not
be congruent with the identities of present-day individuals.
In the words of Manasseh Bueno Barzillai Azevedo da Costa, the protagonist of The King of Schnorrers~\cite{zangwill}, ``Never before have I sat at the table of a Tedesco [Ashkenazi] -- but you -- you are a man after my own heart.  Your soul is a son of Spain.''

\section*{Materials and Methods}
The dynamics of genealogical identity can be derived as follows.
Consider a population of individuals each with state $s_i\in \{(1,1), (2,0), (2,1)\}$.  Denote $\ps_{1,1}$, $\ps_{2,0}$ and $\ps_{2,1}$ as the 
population wide probability of a randomly selected individual to have
the state denoted in the subscript, e.g., $\ps_{1,1}\equiv \frac{1}{N}\sum_{i=1}^N \delta_{s_i,(1,1)}$.  
For convenience, define the time-varying fraction of type-2 individuals
with no type-1 ancestors as $\eps=\ps_{2,0}/\left(\ps_{2,0}+\ps_{2,1}\right)$.
Initially,
$\ps_{1,1}(0)=p_0$, $\ps_{2,0}(0)=1-p_0$, and $\ps_{2,1}(0)=0$.
In generation $g=1$, $\ps_{2,0}$ is equal to the
fraction of individuals both of whom had type-2 parents, i.e.,
$\ps_{2,0}(1)=(1-p)(1-\alpha p)$, and $\eps(1)=(1-\alpha p_0)$.
In subsequent generations,
some type-2 individuals themselves have type-1 ancestors.  
Of the type-2 individuals born to parents who are both type-2, only
a fraction, $\eps^2$, are a product of reproduction involving
parents with no type-1 ancestors.  
Hence, $\ps_{2,0}(2)=(1-p(1))(1-\alpha p(1))\eps(1)^2$.  
Recall that the genealogical dynamics
modeled here preserves the fraction of typed individuals, on average,
a result leveraged by assuming that $p(g)=p_0$.  
Therefore, the predicted discrete-time population dynamics of
the system can be written, for $g>0$ as:
\begin{eqnarray*}
\ps_{1,1}(g+1)&=&p_0, \\
\ps_{2,0}(g+1)&=&\left[(1-p_0)(1-\alpha p_0)\right]\eps(g)^2, \\
\ps_{2,1}(g+1)&=&\alpha p_0\left(1-p_0\right)+\left[(1-p_0)(1-\alpha p_0)\right](1-\eps(g)^2)
\end{eqnarray*}
In this framework,
\begin{equation*}
\ps_{2,0}(g+1)=\left[(1-p_0)(1-\alpha p_0)\right]\left[\frac{\ps_{2,0}(g)}{\ps_{2,0}(g)+\ps_{2,1}(g)}\right]^2 
\label{eq:ps20}
\end{equation*}
But, recalling that $\ps_{2,0}+\ps_{2,0}=1-p_0$, this can be simplified as:
\begin{equation}
\ps_{2,0}(g+1)=\frac{(1-\alpha p_0)}{1-p_0}\ps_{2,0}(g)^2.
\label{eq:recurse}
\end{equation}
Eq.\ref{eq:recurse} can be solved recursively, implying
that the subpopulation of type-2 individuals with no type-1 ancestors
declines superexponentially so long as $0<\alpha\leq 1$:
$\ps_{2,0}(g)=(1-p_0)\left(1-\alpha p_0\right)^{2^g-1}$.
Finally, the fraction of individuals with at least one type-1 ancestor
is $q(g)=1-\ps_{2,0}(g)$, as stated in the main text.

\section*{Acknowledgments}
The author thanks M.~Cortez, J.~Dushoff, D.~Goldman, M.~Goodisman, S.~Levin, S.~Manrubia, S.~Strogatz, R.~Weitz-Shapiro,
and A.~Zangwill for many helpful comments and suggestions.  
Joshua S.~Weitz holds a Career Award at the Scientific Interface from the 
Burroughs Wellcome Fund.

\bibliography{sjews_weitz.bib}

\end{document}